**Demagnetization factor dependence of energy of thick ferromagnetic films**

P. Samarasekara

Department of Physics, University of Peradeniya, Peradeniya, Sri Lanka.

**Abstract**

Second order perturbed Heisenberg Hamiltonian was employed to study the variation of energy of ferromagnetic thick films with demagnetization factor. Under the influence of demagnetization factor given by $\frac{N_d}{\mu_0 \omega}$=6.6, the sc(001) film with 10000 layers can be easily oriented in 0.6 radians direction for the values of energy parameters used in this report. Easy direction of thick fcc(001) film with 10000 layers was determined as 0.66 radians, when the demagnetization factor is given by $\frac{N_d}{\mu_0 \omega}$=2.6. The energy of sc(001)thick film is larger than that of fcc(001) thick film. But the energy curve of fcc(001) thick film is smoother than that of sc(001).

**1. Introduction:**

Ferromagnetic ultra-thin films have been previously investigated using classical model of Heisenberg Hamiltonian with limited number of terms [1]. As it is difficult to understanding the behavior of exchange anisotropy and its applications in magnetic sensors and media technology, exchange anisotropy has been extensively investigated in recent past [2]. Ferromagnetic films are thoroughly studied nowadays, due to their potential applications in magnetic memory devices and microwave devices. Magnetic properties of ferromagnetic thin films have been investigated using Bloch spin wave theory [3]. Although the magnetization of some thin films is oriented in the plane of the film due to dipole interaction, the out of plane orientation is preferred at the surface due to the broken symmetry of uniaxial anisotropy energy. Also two dimensional Heisenberg model has been used to explain the magnetic anisotropy in the presence of dipole interaction [4]. Magnetic properties of ferromagnetic thin films with alternating super layers have been studied using Ising model [5].

Ultra-thin ferromagnetic films with two and three layers have been studied using Heisenberg Hamiltonian with second order perturbation by us [6]. The angle between easy and hard directions was found to be $90^0$ for sc(001), fcc(001) and bcc(001) ferromagnetic lattices. Variable and constant second and fourth order anisotropy constants have been considered for these calculations [6]. The energy of oriented thick ferromagnetic films has been calculated for continuous and discrete variation for thickness by us [7]. Thick films of bcc(001) ferromagnetic lattice up to 10000 layers have been considered for these simulations. Also the solution of classical Heisenberg Hamiltonian with second order perturbation was found for thick ferromagnetic films up to 10000 layers [8]. Stress induced anisotropy is crucial in magnetic films [9, 11]. The unperturbed, 2$^{nd}$ order perturbed and 3$^{rd}$ order perturbed energy of spinel ferrite and ferromagnetic films was determined by us [10, 12, 13, 14, 15, 16, 17].

**2. Model:**

Heisenberg Hamiltonian of any ferromagnetic film can be given as following [6-8].

$$H = -\frac{J}{2}\sum_{m,n}\vec{S}_m \cdot \vec{S}_n + \frac{\omega}{2}\sum_{m \neq n}\left(\frac{\vec{S}_m \cdot \vec{S}_n}{r_{mn}^3} - \frac{3(\vec{S}_m \cdot \vec{r}_{mn})(\vec{r}_{mn} \cdot \vec{S}_n)}{r_{mn}^5}\right) - \sum_m D_{\lambda_m}^{(2)}(S_m^z)^2 - \sum_m D_{\lambda_m}^{(4)}(S_m^z)^4$$

$$-\sum_{m,n}[\vec{H} - (N_d \vec{S}_n / \mu_0)] \cdot \vec{S}_m - \sum_m K_s \sin 2\theta_m ,$$

where m (or n), N, J, $Z_{|m-n|}$, $\Phi_{|m-n|}$, $\omega$, $\theta_m$ (or $\theta_n$), $D_m^{(2)}$, $D_m^{(4)}$, $H_{in}$, $H_{out}$, $N_d$, $K_s$ are indices of layers, total number of layers, spin exchange interaction, number of nearest spin neighbors, constants arisen from partial summation of dipole interaction, strength of long range dipole interaction, azimuthal angles of spins, second order anisotropy, fourth order anisotropy, in plane applied field, out of plane applied field, demagnetization factor and the stress induced anisotropy factor, respectively.

Beginning from above equation, following equation has been proven in one of our early report [8].

$$E(\theta) = -\frac{J}{2}[NZ_0 + 2(N-1)Z_1] + \{N\Phi_0 + 2(N-1)\Phi_1\}(\frac{\omega}{8} + \frac{3\omega}{8}\cos 2\theta)$$

$$- N(\cos^2 \theta D_m^{(2)} + \cos^4 \theta D_m^{(4)} + H_{in}\sin\theta + H_{out}\cos\theta - \frac{N_d}{\mu_0} + K_s \sin 2\theta)$$

$$- \frac{[-\frac{3\omega}{4}(\Phi_0 + 2\Phi_1) + D_m^{(2)} + 2D_m^{(4)}\cos^2\theta]^2 (N-2)\sin^2 2\theta}{2C_{22}}$$

$$- \frac{1}{C_{11}}[-\frac{3\omega}{4}(\Phi_0 + \Phi_1) + D_m^{(2)} + 2D_m^{(4)}\cos^2\theta]^2 \sin^2 2\theta \quad \textbf{(1)}$$

Here $C_{11} = JZ_1 - \frac{\omega}{4}\Phi_1(1+3\cos 2\theta) - 2(\sin^2\theta - \cos^2\theta)D_m^{(2)}$

$+ 4\cos^2\theta(\cos^2\theta - 3\sin^2\theta)D_m^{(4)} + H_{in}\sin\theta + H_{out}\cos\theta - \frac{2N_d}{\mu_0} + 4K_s\sin 2\theta$

$C_{22} = 2JZ_1 - \frac{\omega}{2}\Phi_1(1+3\cos 2\theta) - 2(\sin^2\theta - \cos^2\theta)D_m^{(2)}$

$+ 4\cos^2\theta(\cos^2\theta - 3\sin^2\theta)D_m^{(4)} + H_{in}\sin\theta + H_{out}\cos\theta - \frac{2N_d}{\mu_0} + 4K_s\sin 2\theta$

All these simulation will be carried out for

$\frac{J}{\omega} = \frac{D_m^{(2)}}{\omega} = \frac{H_{in}}{\omega} = \frac{H_{out}}{\omega} = \frac{K_s}{\omega} = 10$ and $\frac{D_m^{(4)}}{\omega} = 5$.

**3. Results and discussion:**

For sc(001) lattice, $Z_0=4$, $Z_1=1$, $\Phi_0=9.0336$ and $\Phi_1=-0.3275$ [1], 3-D graph of energy versus angle and $\frac{N_d}{\mu_0\omega}$ is given in figure 1 for N=10000. Several energy minimums can be observed at different values of angles and demagnetization factors indicating that these are the easy directions. At $\frac{N_d}{\mu_0\omega}=6.6$, one easy direction can be observed at this 3-D plot. The angle corresponding to this $\frac{N_d}{\mu_0\omega}=6.6$ can be found using energy curve given in figure 2. Angles corresponding to energy minimum and maximum were found to be 0.6 and 2.76 radians, respectively.

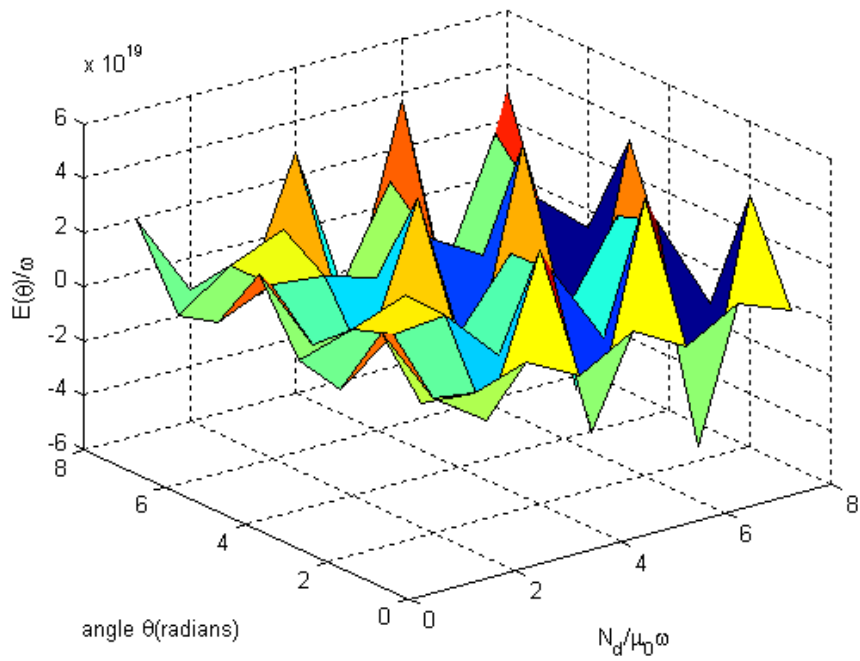

Figure 1: 3-D plot of energy versus angle and $\dfrac{N_d}{\mu_0 \omega}$ for sc(001) thick ferromagnetic film with 10000 layers

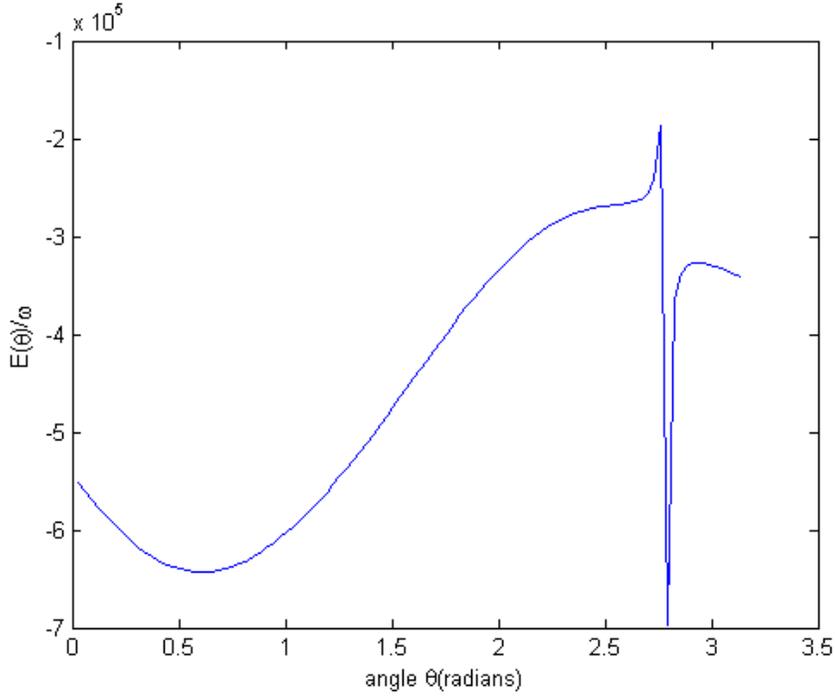

Figure 2: Energy curve of sc(001) film at $\frac{N_d}{\mu_0 \omega}$=6.6 with 10000 layers

For fcc(001) lattice with $Z_0$=4, $Z_1$=4, $\Phi_0$=9.0336, and $\Phi_1$=1.4294 [1], the 3-D plot of energy versus $\frac{N_d}{\mu_0 \omega}$ and angle is given in figure 3 for N=10000. Energy minimums and maximums observed in these graphs are corresponding to easy and hard directions of thick film. For example one easy direction can be found at $\frac{N_d}{\mu_0 \omega}$=2.6. The angle corresponding to this easy direction can be found from figure 4. This energy curve is smoother than the energy curve obtained in previous case. The energy minimum and maximum can be observed at 0.66 and 2.55 radians, respectively.

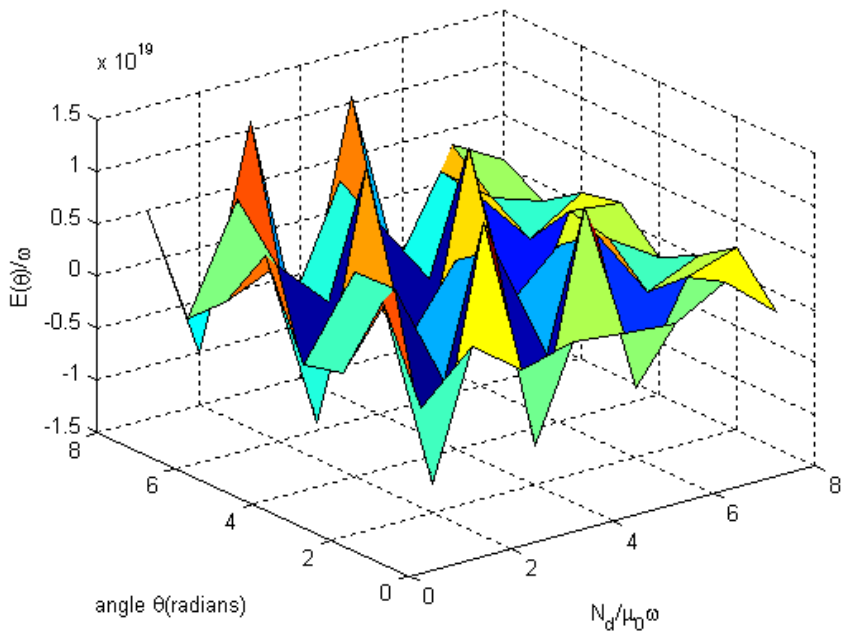

Figure 3: 3-D graph of energy versus angle and $\dfrac{N_d}{\mu_0 \omega}$ for fcc(001) thick ferromagnetic film with 10000 layers

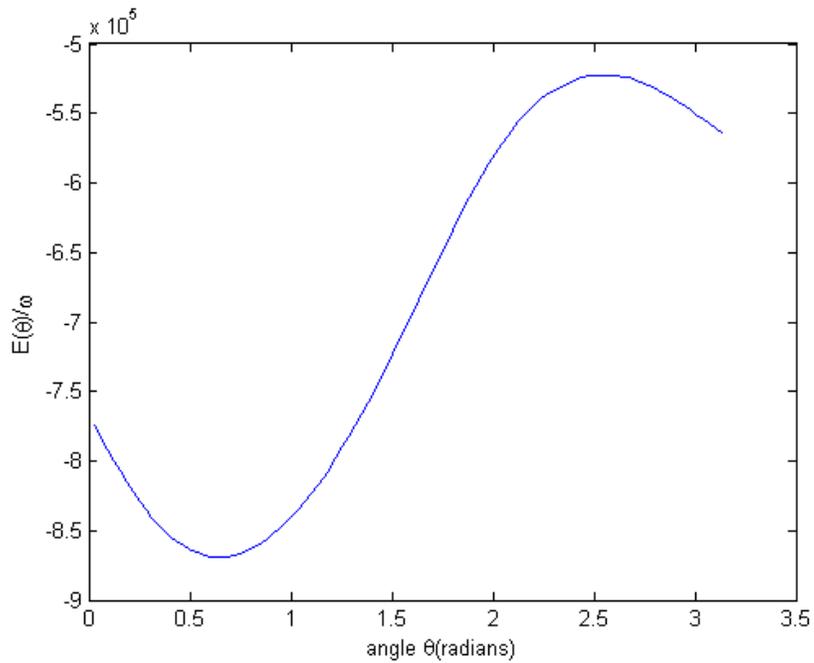

Figure 4: Graph of energy versus angle for fcc(001) film at $\dfrac{N_d}{\mu_0 \omega}=2.6$

## 3. Conclusion:

Many easy directions can be observed at many values of demagnetization factors for sc(001) and fcc(001) thick ferromagnetic films with 10000 layers. The sc(001) film with 10000 layers can be easily oriented in direction given by 0.6 radians under influence of demagnetization factor $\frac{N_d}{\mu_0 \omega}$=6.6. When the demagnetization factor is $\frac{N_d}{\mu_0 \omega}$=2.6, easy direction of thick fcc(001) film with 10000 layers was found to be 0.66 radians. According to 3-D plots, the energy of sc(001) thick film is larger than that of fcc(001) thick film. According to 2-D plots, the energy curve of fcc(001) thick film is smoother than that of sc(001). Although this simulation was carried out for some selected values of energy parameters, this same simulation can be performed for any values of energy parameters.